\documentclass[galaxies,article,accept,oneauthor,LaTeX and dvi2pdf]{Definitions/mdpi} 
\firstpage{1} 
\pubvolume{8}
\issuenum{3}
\articlenumber{62}
\pubyear{2020}
\copyrightyear{2020}
\history{Received: 17 July 2020; Accepted: 20 August 2020; Published: date}

\usepackage{subcaption}
\usepackage[countmax]{subfloat}
\Title{X-Ray Spectral Evolution of High Energy Peaked~Blazars}

\Author{Haritma Gaur}

\AuthorNames{Haritma Gaur}

\address[1]{Aryabhatta Research Institute of Observational Sciences (ARIES), Manora Peak, Nainital 263002, India; harry.gaur31@gmail.com\\}

\abstract{The synchrotron hump of the high energy peaked blazars generally lies in the 0.1--10 keV range and such sources show extreme flux and spectral variability
in X-ray bands. Various spectral studies showed that the X-ray spectra {of high energy peaked blazars} are curved and better described by the log-parabolic model.
{The curvature is attributed to the energy dependent statistical acceleration mechanism}. In this work, we  review the X-ray 
spectral studies of high energy peaked blazars. It is found that the log-parabolic model well describes the spectra in a wide energy interval around the peak.
The log-parabolic model provides the possibility of investigating the correlation between the spectral parameters derived from it. Therefore, we compiled the studies of 
correlations between the various parameters derived from the log-parabolic model and their implications to describe the variability mechanism of blazars.
}

\keyword{BL Lacertae objects: general; galaxies: active; galaxies: jets; radiation mechanisms: non-thermal; X-rays: galaxies}

\begin{document}
%------------------------------------------------------ Introduction -------------------------------------------------------------
\section{Introduction} \label{sec:intro}
Multiwavelength observations of blazars indicate that they belong to AGNs (active galactic nuclei) with relativistic jets pointing close to the line of sight of observer. 
Blazars show rapid flux and spectral variability across the whole electromagnetic (EM) spectrum with the emission being strongly polarized. 
Rapid variability is consistent with the relativistic beaming model, i.e., the bulk relativistic motion of the jet plasma at small angles relative to our line of sight, which gives
 rise to a strong amplification and rapid variability in the observer's frame. 
BL Lacertae objects and flat spectrum radio quasars (FSRQs) are collectively known as blazars. 
The radiation from blazars is predominately non-thermal with spectral energy distribution (SED) characterized by two broad spectral humps 
\citep{Ghisellini et al. (1997),Fossati et al. (1998)}. The first hump is generally attributed to synchrotron emission from relativistic electrons in the jet and 
it lies from radio to the UV or X-rays. A higher frequency hump can be either due to Compton scattering of lower-frequency radiation by the same relativistic electrons 
(synchrotron self compton; SSC) or due to external photons in the broad line region (BLR); torus, etc. (external Compton, EC), i.e.,  
(leptonic models, i.e., \citep{Krawczynski (2004)}). It could also be due to interactions of ultra-relativistic protons in the jet (hadronic models), either via proton synchrotron 
radiation \citep{Mucke et al. (2003)} or via secondary emission from a photo-pion and photo-pair production process (\citep{Bottcher (2007)}; and references therein).  

Depending on the peak frequency of their synchrotron emission, blazars are classified into three sub-classes.
LSPs (low-synchrotron-peaked blazars), consisting predominantly of LBLs (low energy blazars) and defined by a peak of their synchrotron component 
at $\nu_{sy} <$ 10$^{14}$ Hz; ISPs (intermediate-synchrotron-peaked blazars) or IBLs, consisting mostly of intermediate blazars, defined by
10$^{14}$ Hz $< \nu_{sy} <$ 10$^{15}$ Hz; and HSPs (high-synchrotron-peaked blazars), all of which are HBLs
(blue or high energy or X-ray selected blazars) and which are defined through $\nu_{sy} >$ 10$^{15}$~Hz \citep{Abdo (2010)}).
The~high-energy component of the spectral energy distribution (SED) of blazars extends up to $\gamma-$rays,
peaking at GeV energies in LSPs and at TeV energies in HSPs. 

Several HBLs have been detected at TeV energies and have shown
strong X-ray flux variability over diverse time-scales ranging between minutes to years (e.g., \citep{Brinkmann (2005),Zhang (2005),Zhang (2008),Gaur (2010),Kapanadze (2014),Kapanadze (2020)} and references therein). 
The~variability amplitude was found to be correlated with energy, as
the hardest synchrotron radiation is produced by the most energetic electrons with the smallest cooling time-scales (e.g.,
\citep{Zhang (2005),Gliozzi (2006)}). Their X-ray spectra are generally characterized by a soft ($\Gamma$ > 2) convex shape
or by continuously downward curved shape (e.g., \citep{Perlman (2005),Zhang (2008)}), and can originate from an energy dependent particle acceleration 
mechanism with suitable cooling timescales (e.g., \citep{Massaro 2004}). Hence, detailed study of the timing and spectral behavior
of HSPs reveals the most plausible acceleration mechanism which further allows one to draw conclusions about the physical properties of the jet emission
region.  
 
The availability of extensive X-ray observations of blazars enable precise studies of their spectral studies. 
It is well known that the spectra and SEDs of HSPs often appear to be intrinsically curved (\citep{Landau (1986),Massaro 2004,Massaro (2008)})
and the log-parabola model is often invoked to fit the SED in a wide energy interval around the peak (even though in few cases the power law model also provides an acceptable fit).
{This~law is not just a mathematical description of curved spectra, but it can be explained by means of some physical processes. } Reference
{\citep{Massaro 2004}} also showed that a log-parabolic synchrotron spectrum can be obtained by a relativistic electron population having a
similar energy distribution. Curved spectra were related to the radiative ageing of the emitting electrons that have a single power
law injection spectrum~\citep{Massaro 2004}. Reference~{\citep{Kardashev (1962)}}~predicts~a change in the electron spectral index by unity (0.5 for the synchrotron spectrum) around a 
break energy, whose value decreases with time. Probably for this reason, curved spectra have been modeled by means of other complex functions, such as double broken power law or two power laws.    

In this paper, we review the advantages of using the log-parabola model (over power law) to fit curved spectra of HSPs in X-ray bands, {as one can obtain the integral energy 
distribution of the accelerated particles to be a log-parabolic law under a reasonable hypothesis (i.e., \citep{Massaro 2004}). 
Additionally, we summarize the results of correlations between the spectral parameters derived from the log-parabolic model and the implications of these correlations on the particle
acceleration in the jet of HSPs. }

% the study of the relationship between the parameters of synchrotron spectra and those of the electrons. 
%They are generally approximated well by simple power behaviors and that they provide useful information on the electron spectrum.

\section{Models Used to Fit the X-Ray Spectra of Blazars}

The X-ray spectra of blazars are generally fit with following models:
 \begin{enumerate}
 \item A power law model defined by $k~E^{\Gamma}$ with fixed galactic absorption. It is characterized by a normalization $k$ and spectral index $\Gamma$.
In most of the HSPs, X-ray spectra are not well fitted with a power law model and unacceptable values of $\chi^{2}$ are obtained.
{The X-ray spectra of HSPs are remarkably curved}; hence, this model is not adequate to describe their spectra.

{Other analytical models such as a broken power law or a power law with an exponential cut-off (PL plus EC model) are also used to fit the spectra of such sources.
However, the $\chi^{2}$ values are generally be large to be acceptable, indicating that the spectral curvature is milder than an exponential \citep{Massaro (2008)}. } 

\item A logarithmic parabola model defined by $k~E^{-(a+b log(E/E_{1})}$ (e.g., \citep{Landau (1986),Massaro 2004,Massaro (2008)}) with fixed galactic absorption.
 It is characterized by a normalization $k$, spectral index $a$, transition energy $E_{1}$, and curvature parameter $b$. Or the alternative SED representation
\begin{equation}
 S(E) = S_p~10^{-b~log^2(E/E_{p})}~
 \end{equation}
 with $S_p=E_{p}^{2}\, F(E_p)$.
 The values of the parameters $E_p$ (the location of the SED energy peak), $S_p$ (the peak height), and $b$ (the curvature parameter) can be estimated independently 
in the fitting procedure (\citep{Tramacere (2007a)}).

The rest frame energy peak $E_p^*$ is given by
 \begin{equation}
 E_p^* = (1+z)~E_p.
 \end{equation}

 Additionally, the value of $S_p$ is proportional to the bolometric emitted flux, the rest frame powers of BL Lacs in terms of
 the isotropic luminosity peak energy $L_p^*$:
 \begin{equation}
 L_p^* \simeq 4\pi D_L^2 S_p.
 \end{equation}
 where  $D_L$ is the luminosity distance of the blazar.
% \begin{equation}
	% D_L =  \frac{c}{H_0} (1+z) \int_{0}^{z}\frac{dz}{\sqrt{\Omega_M (1+z)^3+\Omega_{\Lambda}}},
% \end{equation}
% using a flat cosmology with $H_0=72$ km/(s Mpc), $\Omega_{M}=0.27$ and
% $\Omega_{\Lambda}=0.73$ (see Spergel et al., 2007).

 \end{enumerate}
 
In the next section, we  summarize the correlations existing between the various parameters derived from the log-parabolic model and their implications in 
predicting the variability mechanism in~blazars.
 
%%%%%%%%%%%%%%%%%%%%%%%%%%%%%%%%%%%%%%%%%%
\section{Correlation between Different Parameters}
\unskip
\subsection{Correlation between $S_{p}$ and $E_{p}$}

As we know that the synchrotron emission mechanism is the dominant component in the X-ray emission of HSPs, a correlation between  $S_{p}$ and $E_{p}$ indicates
the driver of the spectral changes in X-ray band. From \citep{Tramacere (2007a)}, the synchrotron SED is scaled as 
$S \propto N \gamma^{3} B^{2} \delta^{4}$ at the energies $E \propto \gamma^{2} B \delta$  where N is the total emitted number, B is the magnetic field,
$\gamma$m c$^{2}$ is the total electron energy, and $\delta$ is the beaming factor. Hence, the dependence of $S_{p}$ on $E_{p}$ is a power law; i.e., $S_{p}  \propto E_{p}^{\alpha}$.
Now, $\alpha$=1.5 when the spectral changes are dominated by variations of the electron average energy; $\alpha$ = 2 arises from the changes in the magnetic field;
$\alpha$ = 4 arises when changes in the beaming factor dominate and $\alpha$ = $\infty$ arises for changes in the number of emitting particles (\citep{Tramacere (2007a)}). 
 
Reference {\citep{Tramacere (2007a)}} performed spectral analysis of Mrk 421 using the archived observations of {\it ASCA}, {\it BeppoSAX}, {\it RXTE}, and {\it XMM--Newton} satellites
and a positive correlation was found between the position and height of the SED peak with $\alpha$ = 1.5 and 2, which is
compatible with a combined effect of variations of B (corresponding to $\alpha$ = 2) and of a rescaling of $\Gamma$ (corresponding to $\alpha$ = 1.5).
Reference \citep{Tramacere (2007b)} presented the spectral analysis of a sample of HSPs performed by the XRT and UVOT detectors onboard the {\it Swift} satellite
in 2005 and found that the spectra of most of the HSP blazars are curved. Reference {\citep{Massaro (2008)}} performed spectral analysis of a sample of 15 blazars
with the {\it BeppoSAX, XMM--Newton and Swift} satellites and found a positive correlation between $S_{p}$ and $E_{p}$ in five of the blazars.
 Reference {\citep{Kapanadze (2020)}} presented {\it Swift} observations of Mrk 421 during 2015--2018 and found a relation $S_{p} \propto E_{p}^{\alpha}$ with
$\alpha$ $\sim$ 0.6 which indicates the existence of stochastic acceleration scenario in blazar emission.
 %Reference [21] are missing. Please renumber the references so they appear in sequential numerical order.
 
 {Recently, reference \citep{Yijun-Shifu (2019)} studied a sample of 14 blazars from Swift and RXTE and found that nine out of fourteen blazars showed a positive correlation between $L_{p}$ and $E_{p}$ with $\alpha$ $\sim$ 1; thus,
their spectral variations are mainly caused by the changes of the electron energy while the total electron number may remain constant.
The rest of the sources showed a negative or no correlation  between $L_{p}$ and $E_{p}$, and therefore  aforementioned mechanisms did not apply.
One source, 1ES 1959 + 650, had two different $L_{p}$ and $E_{p}$ relations in 2002 and 2016 which indicates that the causes of spectral
variations likely changed during these two periods. }

\subsection{Correlation between $E_{p}$ and $b$ (Curvature Parameter)}

Correlation studies between $E_{p}$  and $b$ indicate the possible signatures of the electron statistical/stochastic acceleration processes in the spectral evolution 
of the synchrotron component. In~this acceleration scenario, the probability for a particle to be further accelerated decreases with energy (i.e., \citep{Massaro (2006)}).
The electron energy distribution is curved into a log-parabolic shape in these two mechanism, and its curvature $r$ is related to the fractional acceleration gain 
$\epsilon$ as follows:
$r \propto  1/(log \epsilon)$.

{In the statistical acceleration process, electron energy distribution follows a log-parabolic law and $E_{p}$ and $b$ follow the relation 
log $E_{p}$ $\sim$ Const. + 2/(5b), given the assumption of $b=r/4$~\citep{Chen (2014)}. The electron energy distribution follows a log normal law for the
fluctuations of the fractional acceleration gain process. In this case, energy gain fluctuations are a random variable around the systematic energy gain \citep{Tramacere (2011)},
and $E_{p}$ and $b$ follow the relation log $E_{p}$ $\sim$ Const. + 3/(10b) \citep{Chen (2014)}.
In the stochastic acceleration process,  a momentum diffusion term is included in the kinetic equation, which leads to energy gain fluctuations in
the diffusive shock acceleration process \citep{Tramacere (2011)}. In this case,  $E_{p}$ and $b$ follow the relation
of log $E_{p}$ $\sim$ Const. + 1/(2b) given b = r/4 \citep{Chen (2014)}.} Hence, an anti-correlation is expected between $E_{p}$ and $\beta$. 
 
Reference {\citep{Tramacere (2007a)}} found an inverse correlation between $E_{p}$ and $b$ which may be interpreted in the frame work of acceleration processes of the 
emitting electrons. Similarly, {\citep{Massaro (2008)}} found a negative correlation between $E_{p}$ and $b$ for a sample of five blazars.
Reference {\citep{Sinha (2015)}} analyzed the observations of Mrk 421 of 2013 from {\it Swift} and {\it NuSTAR} satellites and did not find any significant correlation between
$E_{p}$ and~$b$. Reference~{\citep{Kapanadze (2020)}}~found~a negative correlation between $E_{p}$ and $b$, which was explained by the stochastic acceleration of X-ray 
emitting electrons by the magnetic turbulence. {Reference {\citep{Chen (2014)}} fitted the broadband SEDs of 48 blazars with two log-parabolic models and found a negative correlation between
$E_{p}$ and $b$ for the whole sample. Reference~{\citep{Yijun-Shifu (2019)}}~also studied 14 HSPs and found that seven of the sources showed negative correlations between $E_{p}$ and $b$
with constant values consistent with either the energy dependent acceleration probability scenario or the stochastic accerelation process. For the other seven sources, 
no correlation exists between $E_{p}$ and $b$, and hence they cannot be explained by any above-mentioned mechanism.}

\subsection{Correlation between $a$ and $b$}

Due to the particle acceleration mechanism described above, there exists a positive correlation between $a$ and $b$ (\citep{Massaro 2004}).
The probability of the particle's acceleration is lower when its energy increases. The energy spectrum is given as follows:

$N(\gamma)\sim{\gamma/\gamma_0}^{-s-r\log{\gamma/\gamma_0}}$, with a linear relationship between the spectral index and curvature terms
(\emph{s} and \emph{r}, respectively) $s=-r(2/q)\log{g/\gamma_0}-(q-2)/2$. The~synchrotron emission
produced by this distribution is given as follows:
\begin{equation}
P_S (\nu)\propto (\nu/\nu_0)^{-(a+b\log(\nu/\nu_0))},
\end{equation}
with $a=(s-1)/2$ and $b=r/4$  (\citep{Massaro 2004}). 

Reference {\citep{Massaro 2004}} found a positive correlation between these two parameters for their spectral studies of Mrk 421. However, reference {\citep{Sinha (2015)}} did not find any significant correlation
{between these two parameters} for the same source {during a strong flare in 2013}. References 
{\citep{Kapandze (2017a),Kapanadze (2018),Kapanadze (2020)}} {also studied Mrk 421 and Mrk 501 during flaring states} and
 found a positive correlation between these two parameters {in most of the epochs }. However, they found weaker correlation during 2009--2012 for the observations
of Mrk~421~(\mbox{\citep{Kapanadze (2016),Kapanadze (2017b)}}).  Reference~{\citep{Katarzynski (2006)}}~performed~simulations which revealed that the electrons which are accelerated at the
shock front can continue gaining energy via the stochastic mechanism into the shock acceleration region and repeat the acceleration cycle. Since the linear
correlation is not expected within the stochastic acceleration mechanism, it could weaken the aforementioned correlation and therefore soometimes weaker correlation is found between 
these two parameters.  

\subsection{Correlation between $E_{p}$ and flux}

It was found that blazars exhibit strong spectral variability along with the flux variability (\citep{Massaro 2004,Gaur (2010)}). The synchrotron SED of HSP 
blazars shows a positive correlation with respect to flux; i.e., it shifted by several keV to higher energies during high flux states and moved back as the flux decreases. 
The~spectral shape generally follows the "harder-when-brighter" trend (\citep{Kapanadze (2020),Pandey  (2017),Paliya (2015),Balokovic (2016)}).
 Reference {\citep{Kapanadze (2020)}} found extreme variability in the synchrotron SED peak which varies between $E_{p}$ $<$ 0.1 keV (at low flux states) and \mbox{$E_{p}$
$>$15 keV} at high flux states. A similar relationship between flux and $E_{p}$ was found by reference \citep{Sinha (2015)}; i.e., the spectral index hardenss and the
peak of the spectrum shifts to higher energies.

\section{Conclusions}

HSP blazars are extensively variable in X-ray bands due to their synchrotron emission hump peak in the 0.1--10 keV range. This flux variability is 
often associated with spectral variability. Interpretation of spectral variability is not very simple, as it involves many non-stable processes, such as particle 
acceleration, injection, cooling, and escape, which
contain a number of unknown physical parameters. Hence, the curved spectra of such sources are not well described by simple power laws, but they are intrinsically
curved, and hence the log-parabolic spectral model better describes their spectra. Spectral curvature can be interpreted in terms of radiation cooling of high energy
electron population, injected with a power law spectrum. Flux and spectral variability in blazars can be accounted by several processes, e.g., particle acceleration,
injection, diffusion through the emitting volume, radiative cooling, and particle escape. They all are associated with different time-scales and it is not simple to identify 
the dominant process contributing to observed variability. {In case of HSP blazars, in~which the synchrotron emission peaks in X-rays, the radiative cooling time can be very
short. In~the bulk frame of the electrons moving down the relativistic jet, for $\nu$ $\sim$ $10^{16}$ Hz, magnetic field \mbox{B $\sim$ 1} Gauss, and beaming factor $\delta$ 
$\sim$ 10, one can obtain the radiative lifetimes of the order of an hour. Hence, the high energy emission can be maintained only if the acceleration mechanisms/injection are continuously
at work.} Reference {\citep{Massaro 2004}} argued that the log-parabolic model better describes the spectra, 
{as the electron energy distribution itself follows the log-parabolic law for the energy dependent acceleration probability process. Additionally, the acceleration
efficiency of the particles is inversely proportional to their energy.} The parameters derived from the log-parabolic model
show correlations between them {which could indicate the presence of the statistical/stochastic acceleration processes in the X-ray emission of~blazars.}

The important correlations are summarized below:

\begin{enumerate}
\item The height of the SED $S_{p}$ ({or luminosity peak energy $L_{p}$) and $E_{p}$ of the synchrotron peak follows a power law.
Additionally, the power law index indicates the driver of the spectral changes in X-ray band of HSPs.}

 \item Curvature parameter $b$ decreases as $E_{p}$ increases, which indicates that statistical/stochastic acceleration processes are at work for the emitting electrons.

 \item Curvature parameter $b$ and photon index $a$ are also expected to be linearly correlated with respect to each other in the first-order Fermi acceleration. However,
in many observations weaker or absent correlation is found between these two parameters,  which indicates that stochastic (second-order Fermi acceleration)  also exits.

 \item The spectral shape generally shows a "harder-when-brighter" trend. Additionally, $E_{p}$ shifts to higher frequencies as flux increases. 
\end{enumerate}

%%%%%%%%%%%%%%%%%%%%%%%%%%%%%%%%%%%%%%%%%%%
\funding{The author acknowledge the financial support from the Department of Science and Technology, India, through INSPIRE
faculty award IFA17-PH197 at ARIES, Nainital. {I would like to thank referees for their useful~comments.} }
%%%%%%%%%%%%%%%%%%%%%%%%%%%%%%%%%%%%%%%%%%
%%%%%%%%%%%%%%%%%%%%%%%%%%%%%%%%%%%%%%%%%%
\conflictsofinterest{The authors declare no conflict of interest.}

%-------------------------------------------------------- References --------------------------------------------------

\reftitle{References}


\begin{thebibliography}{999}
\providecommand{\natexlab}[1]{#1}

% Reference 1
\bibitem{Ghisellini et al. (1997)}
Ghisellini, G.; Villata, M.; Raiteri, C.M.; Bosio, S.; de Francesco, G.; Latini, G.; Maesano, M.; Massaro,~E.; Montagni, F.; Nesci, R.; et al. Optical-IUE observations of the gamma-ray loud BL Lacertae object S5 0716+714: Data and interpretation. {\em Astron. Astrophys.} {\bf 1997}, {\em 327}, 61–71. [\href{https://academic.oup.com/mnras/article/299/2/433/1019239}{CrossRef}]

\bibitem[Fossati et al. (1998)]{Fossati et al. (1998)}
Fossati, G.; Maraschi, L.; Celotti, A.; Comastri, A.; Ghisellini, G. A unifying view of the spectral energy distributions of blazars. {\em Mon. Not. R. Astron. Soc.} {\bf 1998}, {\em 299}, 433–448. [\href{https://academic.oup.com/mnras/article/299/2/433/1019239}{CrossRef}]

\bibitem[Krawczynski (2004)]{Krawczynski (2004)}
Krawczynski, H. TeV blazars-observations and models. {\em New Astron. Rev.} {\bf 2004}, {\em 48}, 367–373. [\href{https://www.sciencedirect.com/science/article/abs/pii/S138764730300366X?via\%3Dihub}{CrossRef}]

\bibitem[Mucke et al. (2003)]{Mucke et al. (2003)}
Mücke, A.; Protheroe, R. J.; Engel, R.; Rachen, J. P.; Stanev, T. BL Lac objects in the synchrotron proton
blazar model. {\em Astropart. Phys.} {\bf 2003}, {\em 18}, 593–613. [\href{https://www.sciencedirect.com/science/article/abs/pii/S0927650502001858?via\%3Dihub}{CrossRef}]

\bibitem[Bottcher (2007)]{Bottcher (2007)}
B$\ddot{o}$ttcher, M. Modeling the emission processes in blazars. {\em Astrophys. \& Space Sci.} {\bf 2007}, {\em 309}, 95–104.

\bibitem[Abdo (2010)]{Abdo (2010)}
Abdo, A.A.; Ackermann, M.; Agudo, I.; Ajello, M.; Aller, H.D.; Aller, M.F.; Angelakis, E.; Arkharov, A,A.; Axelsson, M.; Bach, U.; et al. The Spectral Energy Distribution of Fermi Bright Blazars. {\em Astrophys. J.} {\bf 2010}, {\em 716}, {\em 30}. [\href{https://iopscience.iop.org/article/10.1088/0004-637X/716/1/30}{CrossRef}]

\bibitem[Brinkmann (2005)] {Brinkmann (2005)}
Brinkmann, W.; Papadakis, I.E.; Raeth, C.; Mimica, P.; Haberl, F. XMM-Newton timing mode observations of Mrk 421. {\em Astron. Astrophys.} {\bf 2005}, {\em 443}, 397--411. [\href{https://www.aanda.org/articles/aa/abs/2005/44/aa2767-05/aa2767-05.html}{CrossRef}]

\bibitem[Zhang (2005)]{Zhang (2005)} 
Zhang, Y.H.; Treves, A.; Celotti, A.; Qin, Y.P.; Bai, J.M. XMM-Newton View of PKS 2155–304: Characterizing the X-Ray Variability Properties with EPIC pn. {\em Astrophys. J.} {\bf 2005}, {\em 629}, 686--699. [\href{https://iopscience.iop.org/article/10.1086/431719}{CrossRef}]
% Reference 12

\bibitem[Zhang (2008)]{Zhang (2008)}
Zhang, Y.H. XMM-Newton Observations of the TeV BL Lacertae Object PKS 2155-304 in 2006: Signature of Inverse Compton X-Ray Emission? {\em Astrophys. J.} {\bf 2008}, {\em 682}, 789--797. [\href{https://iopscience.iop.org/article/10.1086/589493}{CrossRef}]
 
\bibitem[Gaur (2010)]{Gaur (2010)}
Gaur, H.; Gupta, A.C.; Lachowicz, P.; Wiita, P.J. Detection Of Intra-day Variability Timescales of Four High-Energy Peaked Blazars With XMM-Newton. {\em Astrophys. J.} {\bf 2010}, {\em 718}, 279–291. [\href{https://iopscience.iop.org/article/10.1088/0004-637X/718/1/279}{CrossRef}]

\bibitem[Kapanadze (2014)]{Kapanadze (2014)}
Kapanadze, B.; Romano P.; Vercellone, S.; Kapanadze, S. The X-ray behaviour of the high-energy peaked BL Lacertae source PKS 2155-304 in the 0.3-10 keV band. {\em Mon. Not. R. Astron. Soc.} {\bf 2014}, {\em 444}, 1077--1094. [\href{https://academic.oup.com/mnras/article/444/2/1077/1747666}{CrossRef}]

\bibitem[Kapanadze (2020)]{Kapanadze (2020)}
Kapanadze, B. Swift Observations of Mrk 421 in Selected Epochs. III. Extreme X-Ray Timing/Spectral Properties and Multiwavelength Lognormality during 2015 December-2018 April. {\em  Astrophys. J. Suppl. Ser.} {\bf 2020}, {\em 247}, 27K. [\href{https://iopscience.iop.org/article/10.3847/1538-4365/ab6322}{CrossRef}]

% Reference 14
\bibitem[Gliozzi (2006)]{Gliozzi (2006)}
Gliozzi, M.; Sambruna, R.M.; Jung, I.; Krawczynski, H.; Horan, D.; Tavecchio, F. Long-Term X-Ray and TeV Variability of Mrk 501. {\em Astrophys. J.} {\bf 2006}, {\em 646}, 61. [\href{https://iopscience.iop.org/article/10.1086/504700}{CrossRef}]

\bibitem[Perlman (2005)]{Perlman (2005)}
Perlman, E.S.; Madejski, G.; Georganopoulos, M.; Andersson, K.; Daugherty, T.; Krolik, J.H.; Rector, T.; Stocke, J.T.; Koratkar, A.; Wagner, S.; et al. Intrinsic Curvature in the X-Ray Spectra of BL Lacertae Objects. {\em Astrophys. J.} {\bf 2005}, {\em 625}, 727. [\href{https://iopscience.iop.org/article/10.1086/429688}{CrossRef}]

\bibitem[Massaro (2004)]{Massaro 2004}
Massaro, E.; Perri, M.; Giommi, P.; Nesci, R. Log-parabolic spectra and particle acceleration in the BL Lac object Mkn 421: Spectral analysis of the complete BeppoSAX wide band X-ray data set. {\em Astron. Astrophys.} {\bf 2004}, {\em 413}, 489. [\href{https://www.aanda.org/articles/aa/abs/2004/02/aa4010/aa4010.html}{CrossRef}]

\bibitem[Landau (1986)]{Landau (1986)}
 Landau, R.; Golisch, B.; Jones, T.J.; Jones, T.W.; Pedelty, J.; Rudnick, L.; Sitko, M.L.; Kenney, J.; Roellig, T.; Salonen, E.; et al. Active Extragalactic Sources: Nearly Simultaneous Observations from 20 Centimeters to 1400 Angstrom. {\em Astrophys. J.} {\bf 1986}, {\em 308}, 78--92. [\href{https://ui.adsabs.harvard.edu/abs/1986ApJ...308...78L/abstract}{CrossRef}]

\bibitem[Massaro (2008)]{Massaro (2008)}
Massaro, F.; Tramacere, A.; Cavaliere, A.; Perri, M.; Giommi, P. X-ray spectral evolution of TeV BL Lacertae objects: Eleven years of observations with BeppoSAX, XMM-Newton and Swift satellites. {\em Astron. Astrophys.}  {\bf 2008}, {\em 478}, 395. [\href{https://www.aanda.org/articles/aa/abs/2008/05/aa8639-07/aa8639-07.html}{CrossRef}]

\bibitem[Kardashev (1962)]{Kardashev (1962)}
Kardashev, N.S. Nonstationarity of Spectra of Young Sources of Nonthermal Radio Emission. {\em Sov. Astron.} {\bf 1962}, {\em 6}, 317.

\bibitem[Tramacere (2007a)]{Tramacere (2007a)}
Tramacere, A.; Massaro, F.; Cavaliere, A. Signatures of synchrotron emission and of electron acceleration in the X-ray spectra of Mrk 421. {\em Astron. Astrophys.} {\bf 2007}, {\em 466}, 521–529. [\href{https://www.aanda.org/articles/aa/abs/2007/17/aa6723-06/aa6723-06.html}{CrossRef}]

\bibitem[Tramacere (2007b)]{Tramacere (2007b)}
Tramacere, A.; Giommi, P.; Massaro, E.; Perri, M.; Nesci, R.; Colafrancesco, S.; Tagliaferri, G.; Chincarini, G.; Falcone, A.; Burrows, D.N.; et al. SWIFT observations of TeV BL Lacertae objects. {\em Astron. Astrophys.} {\bf 2007}, {\em 467}, 501--508. [\href{https://www.aanda.org/articles/aa/abs/2007/20/aa6226-06/aa6226-06.html}{CrossRef}]

%\bibitem[Rybicki Lightman (1979)]{Rybicki Lightman (1979)}
%Rybicki, G.B.; Lightman, A.P. \emph{Radiative Processes in Astrophysics}; Wiley: New York, NY, USA, 1979.

\bibitem[Yijun-Shifu (2019)]{Yijun-Shifu (2019)}
 Wang, Y.; Zhu, S. X-Ray Spectral Variations of Synchrotron Peak in BL Lacs. {\em Astrophys. J.} {\bf 2019}, {\em 885}, 8W. [\href{https://iopscience.iop.org/article/10.3847/1538-4357/ab4416}{CrossRef}]

\bibitem[Massaro (2006)]{Massaro (2006)}
Massaro, E.; Tramacere, A.; Perri, M.; Giommi, P.; Tosti, G. Log-parabolic spectra and particle acceleration in blazars III. SSC emission in the TeV band from Mkn 501. {\em Astron. Astrophys.} {\bf 2006}, {\em 448}, 861--871. [\href{https://www.aanda.org/articles/aa/abs/2006/12/aa3644-05/aa3644-05.html}{CrossRef}]

\bibitem[Chen (2014)]{Chen (2014)}
Chen, L. Curvature of the spectral energy distributions of blazars. {\em Astrophys. J.} {\bf 2014}, {\em 788}, 179. [\href{https://iopscience.iop.org/article/10.1088/0004-637X/788/2/179}{CrossRef}]

\bibitem[Tramacere (2011)]{Tramacere (2011)}
Tramacere, A.; Massaro, E.; Taylor, A.M. stochastic acceleration and the evolution of spectral distributions in synchro-self-compton sources: A self-consistent modeling of blazars’ flares. {\em Astrophys. J.} {\bf 2011}, {\em 739}, 66. [\href{https://iopscience.iop.org/article/10.1088/0004-637X/739/2/66}{CrossRef}]


\bibitem[Sinha (2015)]{Sinha (2015)}
Sinha, A.; Shukla, A.; Misra, R.; Chitnis, V.R.; Rao, A.R.; Acharya, B.S. Underlying particle spectrum of Mkn 421 during the huge X-ray flare in April 2013. {\em Astron. Astrophys.} {\bf 2015}, {\em 580}, A100. [\href{https://www.aanda.org/articles/aa/abs/2015/08/aa26264-15/aa26264-15.html}{CrossRef}]


\bibitem[Kapandze (2017a)]{Kapandze (2017a)}
Kapanadze, B.; Dorner, D.; Romano, P.; Vercellone, S.; Mannheim, K.; Lindfors, E.; Nilsson, K.; Reinthal,~R.; Takalo, L.; Kapanadze, S.; et al. The prolonged X-ray flaring activity of Mrk 501 in 2014. {\em Mon. Not. R. Astron.~Soc.} {\bf 2017}, {\em 469}, 1655--1672. [\href{https://academic.oup.com/mnras/article/469/2/1655/3738095}{CrossRef}]

\bibitem[Kapanadze (2018)]{Kapanadze (2018)}
Kapanadze, B.; Vercellone, S.; Romano, P.; Hughes, P.; Aller, M.; Aller, H.; Kharshiladze, O.; Kapanadze,~S.; Tabagari, L. Swift Observations of Mrk 421 in Selected Epochs. I. The Spectral and Flux Variability in 2005–2008. {\em Astrophys. J.} {\bf 2018}, {\em 854}, 66. [\href{https://iopscience.iop.org/article/10.3847/1538-4357/aaa75d}{CrossRef}]

\bibitem[Kapanadze (2016)]{Kapanadze (2016)}
Kapanadze, B.; Dorner, D.; Vercellone, S.; Romano, P.; Aller, H.; Aller, M.; Hughes, P.; Reynolds, M.; Kapanadze, S.; Tabagari, L. X-Ray Flaring Activity of Mrk 421 In The First Half Of 2013. {\em Astrophys. J.} {\bf 2016}, {\em 831}, 102. [\href{https://iopscience.iop.org/article/10.3847/0004-637X/831/1/102}{CrossRef}]

\bibitem[Kapanadze (2017b)]{Kapanadze (2017b)}
Kapanadze, B.; Dorner, D.; Romano, P.; Vercellone, S.; Kapanadze, S.; Tabagari, L. Mrk 421 after the Giant X-Ray Outburst in 2013. {\em Astrophys. J.} {\bf 2017}, {\em 848}, 103. [\href{https://iopscience.iop.org/article/10.3847/1538-4357/aa8ea6}{CrossRef}]

\bibitem[Katarzynski (2006)]{Katarzynski (2006)}
Katarzyński, K.; Ghisellini, G.; Mastichiadis, A.; Tavecchio, F.; Maraschi, L. Stochastic particle acceleration and synchrotron self-Compton radiation in TeV blazars. {\em Astron. Astrophys.} {\bf 2006}, {\em 453}, 47. [\href{https://www.aanda.org/articles/aa/abs/2006/25/aa4176-05/aa4176-05.html}{CrossRef}]

\bibitem[Pandey (2017)]{Pandey (2017)}
Pandey, A.; Gupta, A.C.; Wiita, P.J. X-Ray Intraday Variability of Five TeV Blazars with NuSTAR. {\em Astrophys.~J.} {\bf 2017}, {\em 841}, 123. [\href{https://iopscience.iop.org/article/10.3847/1538-4357/aa705e}{CrossRef}]

\bibitem[Paliya (2015)]{Paliya (2015)}
Paliya, V.S.; Böttcher, M.; Diltz, C.; Stalin, C.S.; Sahayanathan, S.; Ravikumar, C.D. The Violent Hard X-Ray Variability of Mrk 421 Observed by NuSTAR in 2013 April. {\em Astrophys. J.} {\bf 2015}, {\em 811}, 143. [\href{https://iopscience.iop.org/article/10.1088/0004-637X/811/2/143}{CrossRef}]


\bibitem[Balokovic (2016)]{Balokovic (2016)}
Baloković, M.; Paneque, D.; Madejski, G.; Furniss, A.; Chiang, J.; Ajello, M.; Alexander, D.M.; Barret,~D.; Blandford, R.D.; Boggs, S.E.; et al. Multiwavelength Study of Quiescent States of Mrk 421 With Unprecedented Hard X-Ray Coverage Provided By NuSTAR in 2013. {\em  Astrophys. J.} {\bf 2016}, {\em 819}, 156. [\href{https://iopscience.iop.org/article/10.3847/0004-637X/819/2/156}{CrossRef}]


\end{thebibliography}
\end{document}